\documentclass[preprint,prl,aps,showpacs,amsmath,amssymb]{revtex4-1}

\usepackage{graphicx}
\usepackage{epstopdf}
\usepackage{braket}
\usepackage[euler]{textgreek}
\usepackage{hyperref}
\usepackage{float}

\begin{document}

\title{Observation of surface superstructure induced by systematic vacancies in the topological Dirac semimetal Cd$_{3}$As$_{2}$}

\author{Christopher J. Butler$^{1,}$\footnote{These authors contributed equally to this work}$^{,}$\footnote{cjbutler@ntu.edu.tw}}
\author{Yi Tseng$^{1,}$\footnotemark[1]}
\author{Cheng-Rong Hsing$^{2}$}
\author{Yu-Mi Wu$^{1}$}
\author{Raman Sankar$^{3,4}$}
\author{Mei-Fang Wang$^{1}$}
\author{Ching-Ming Wei$^{2,}$\footnote{cmw@phys.sinica.edu.tw}}
\author{Fang-Cheng Chou$^{4,5,6}$}
\author{Minn-Tsong Lin$^{1,2,7,}$\footnote{mtlin@phys.ntu.edu.tw}}

\affiliation{$^{1}$Department of Physics, National Taiwan University, Taipei 10617, Taiwan}
\affiliation{$^{2}$Institute of Atomic and Molecular Sciences, Academia Sinica, Taipei 10617, Taiwan}
\affiliation{$^{3}$Institute of Physics, Academia Sinica, Taipei 11529, Taiwan}
\affiliation{$^{4}$Center for Condensed Matter Sciences, National Taiwan University, Taipei 10617, Taiwan}
\affiliation{$^{5}$National Synchrotron Radiation Research Center, Hsinchu 30076, Taiwan}
\affiliation{$^{6}$Taiwan Consortium of Emergent Crystalline Materials (TCECM), Ministry of Science and Technology, Taipei 10622, Taiwan}
\affiliation{$^{7}$Research Center for Applied Sciences, Academia Sinica, Taipei 11529, Taiwan}

\begin{abstract}

The Dirac semimetal phase found in Cd$_{3}$As$_{2}$ is protected by a $C_{4}$ rotational symmetry derived from a corkscrew arrangement of systematic Cd vacancies in its complicated crystal structure. It is therefore surprising that no microscopic observation, direct or indirect, of these systematic vacancies has so far been described. To this end, we revisit the cleaved (112) surface of Cd$_{3}$As$_{2}$ using a combined approach of scanning tunneling microscopy and \textit{ab initio} calculations. We determine the exact position of the (112) plane at which Cd$_{3}$As$_{2}$ naturally cleaves, and describe in detail a structural periodicity found at the reconstructed surface, consistent with that expected to arise from the systematic Cd vacancies. This reconciles the current state of microscopic surface observations with those of crystallographic and theoretical models, and demonstrates that this vacancy superstructure, central to the preservation of the Dirac semimetal phase, survives the cleavage process and retains order at the surface.

\end{abstract}

\maketitle

\newpage

%\section{Introduction}

Dirac and Weyl semimetals host low energy excitations described by the three dimensional (3-D) Dirac/Weyl Hamiltonian \cite{Young2012}. As opposed to the two-dimensional (2-D) linear dispersion found in graphene or at topological insulator surfaces \cite{CastroNeto2009,Hasan2010}, they host analogous `cones' of three-dimensional linear dispersion, emanating from an even number of paired Weyl nodes in the bulk Brillouin zone (BZ). Where time-reversal or inversion symmetries are broken, pairs of isolated Weyl nodes of non-zero chiral charge can exist, giving rise to a Weyl semimetal phase with associated exotic physical properties \cite{Ojanen2013,Hosur2013,Potter2014,Baum2015}. Otherwise, Weyl node pairs are forced to coexist as chiral-charge-neutral Dirac nodes, but protected from gap opening perturbations by additional crystal symmetries. This Dirac semimetal (DSM) phase was first proposed in BiO$_{2}$ and \textit{A}$_{3}$Bi (\textit{A} = Na, K, Rb) \cite{Young2012,Wang2012}, and was realized in Na$_{3}$Bi \cite{Liu2014} and then in the more stable Cd$_{3}$As$_{2}$ \cite{Liu2014a,Neupane2014,Borisenko2014}.

Numerous transport observations have revealed unusual properties associated with the DSM phase in Cd$_{3}$As$_{2}$ \cite{He2014,Feng2015,Liang2014,Cao2015,Li2015,Pan2016,Guo2016}, including those involving surfaces, such as chirality transfer between Weyl nodes through Fermi arc surface states \cite{Moll2016}, and even possible unconventional superconductivity induced at quantum point contact interfaces \cite{Aggarwal2015,Wang2015}.

Dirac nodes are protected by crystal symmetries such as $C_{n}$ rotational, or non-symmorphic glide reflection symmetries. In the case of Cd$_{3}$As$_{2}$, they are protected by a $C_{4}$ rotational symmetry around the [001] axis \cite{Wang2013}, associated with a peculiar arrangement of systematic Cd vacancies in [001]-oriented corkscrews, in an otherwise antifluorite-like structure. This arrangement has been described in detail as part of a thorough bulk structural characterization reported by Ali \textit{et al} \cite{Ali2014}. An apparent discrepancy in the position of the Dirac points (DPs) within the Brillouin zone revolves around these systematic vacancies. The DPs have been located along the (001) direction in angle-resolved photoemission spectroscopy (ARPES) investigations reported by Wang \textit{et al.} \cite{Wang2013}, among others \cite{Jeon2014,Borisenko2014,Ali2014}, but along the (112) direction in work reported by Liu \textit{et al} \cite{Liu2014a}. This discrepancy may be attributable to a possible randomization of the Cd vacancy distribution near the cleaved surface \cite{Jeon2014}. However, it was recently demonstrated that if these Cd vacancies were to be significantly disordered or filled, the symmetry protected band crossing points, and therefore the DSM phase, would be destroyed \cite{Guo2016,Yuan2016}. Even with recent scanning tunneling microscopy (STM) observations \cite{Jeon2014,Ali2014}, it remains that no direct or indirect microscopic observation of these systematic (or random) Cd vacancies has been reported in Cd$_{3}$As$_{2}$, despite their central role in the formation of the DSM phase, and the position of the Dirac nodes within the BZ. Thus a gap potentially exists between the prevailing crystallographic understanding of Cd$_{3}$As$_{2}$ (with its theoretical implications) and existing microscopic results.

In this work we revisit the most easily attainable surface of Cd$_{3}$As$_{2}$, the cleaved (112) surface, using a combined approach of atomically resolved STM measurements and density functional theory (DFT) calculations. We construct an atomistic model of the (112) surface expected according to the bulk structure including systematic Cd vacancies, and show that it is predictive of our microscopic observations. This brings into agreement the crystallographic description and microscopic observations of the systematic Cd vacancy arrangement underpinning the DSM phase in Cd$_{3}$As$_{2}$.

\begin{figure}
\centering
\includegraphics[scale=1]{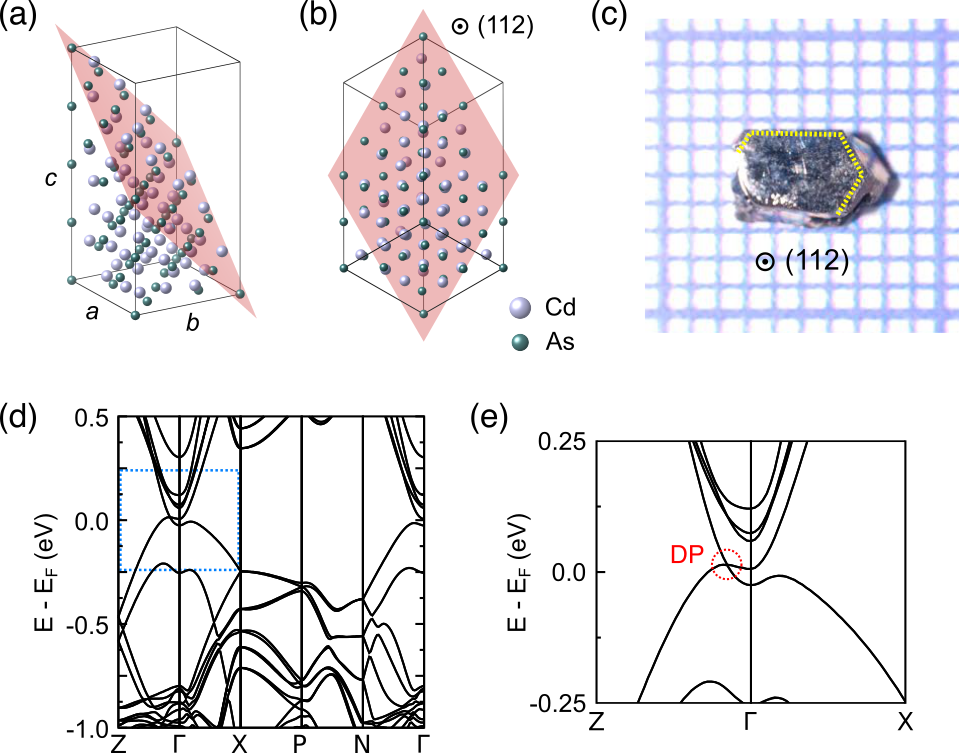}
\caption{\label{fig:1} Bulk crystal and electronic structure of Cd$_{3}$As$_{2}$. (a) A view of the conventional unit cell, with a cut-away plane showing a (112) oriented surface. The pink diamond indicates a (112) plane. (b) The projection of the structure along the (112) axis, showing the pseudo-hexagonal nature of the surface lattice. (c) A photograph of a typical Cd$_{3}$As$_{2}$ platelet as used in this work. (For scale, 1 small square = 1 mm$^{2}$). The primary facet has a hexagonal shape (as indicated by the yellow dashed line) and is identified as a (112) surface. (d) A bulk band structure diagram, and (e) a zoom-in view of the region along \textGamma-Z near the Fermi level, showing the location of the DPs.}
\end{figure}

%\section{Results}

%\subsection{Determination of the (112) cleavage plane}

Figure 1(a) shows the conventional tetragonal Cd$_{3}$As$_{2}$ unit cell reported recently \cite{Sankar2015}, with a portion of the atoms cut away to illustrate the (112) surface. The pink diamond shows one of the (112) planar cuts through the unit cell, parallel to which are the natural cleavage planes. The projection of this structure looking along the normal to the cleavage planes is displayed in Fig. 1(b), showing a pseudo-hexagonal lattice. A photograph of a single crystal Cd$_{3}$As$_{2}$ platelet as used here is shown in Fig. 1(c). Platelets used in this work were grown as described previously \cite{Sankar2015}, and have large shiny primary facets with hexagonal shapes, corresponding to the (112) plane of the crystal. Figure 1(d) shows a bulk electronic band structure diagram calculated for this structure, with high-symmetry axes taken from the structure's smaller primitive cell \cite{Supplement}. Figure 1(e) gives a zoom-in view showing the DP located near the Fermi level between the \textGamma \enspace and Z points of the BZ. 

All band structure and charge density calculations presented in this work were performed in the framework of DFT, using the projector-augmented wave (PAW) method \cite{Blochl1994,Kresse1999} as implemented in the VASP package \cite{Kresse1993,Kresse1996,Kresse1996a}, with the Perdew-Burke-Ernzerhof exchange-correlation functional \cite{Perdew1996} and fully accounting for spin-orbit coupling. Structural optimization was performed using a 2$\times$2$\times$1 Monkhorst-Pack k-point mesh \cite{Pack1977}, and a kinetic energy cutoff of 300 eV. The force on each atom was less than 0.01 eV$\cdot$\AA$^{-1}$.

In order to perform STM measurements on the (112) surface, bulk single crystals of Cd$_{3}$As$_{2}$ were cleaved at room temperature in UHV conditions, before quickly being transferred to an Omicron LT-STM held at 4.5 K. An electrochemically etched tungsten tip was used, and the obtained STM images were processed using the WSxM software package \cite{WSxM}. Figure 2(a) shows the typical step-terrace morphology observed in large-scale STM images taken at the (112) surface of Cd$_{3}$As$_{2}$. Previously, the precise position of the cleavage planes through the crystal had yet to be determined, though it has been suggested by Jeon \textit{et al.}, based on STM surface lattice images, that the cleaved surface terminates with an As atomic layer \cite{Jeon2014}.

\begin{figure} [h]
\centering
\includegraphics[scale=1]{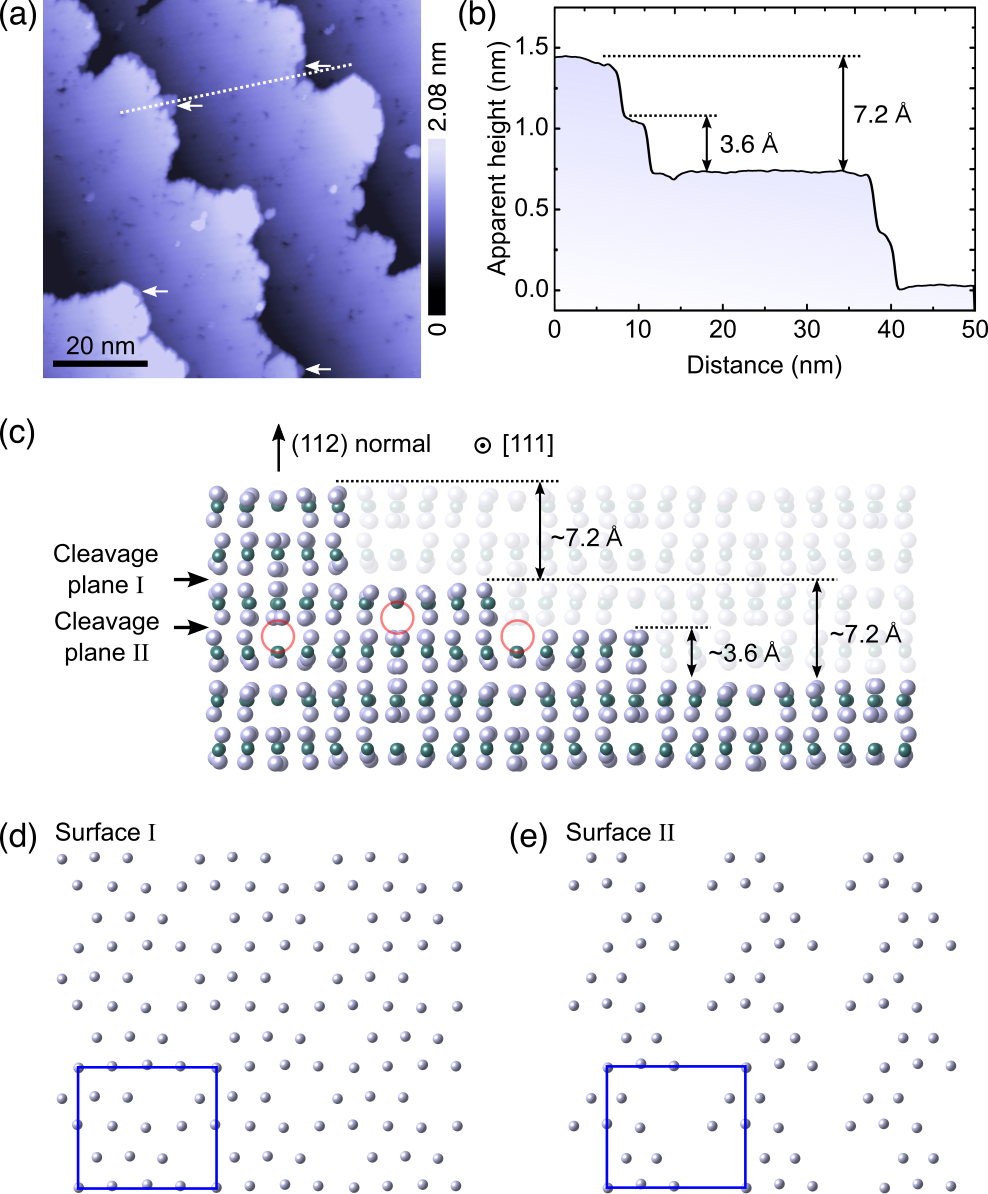}
\caption{\label{fig:2} Interpretation of topographic step heights. (a) A large-scale STM topograph (\textit{V} = 2.0 V, \textit{I} = 0.1 nA), with the topographic height profile, taken along the white dashed line, shown in (b). Small `sub-steps' can be seen in some areas, marked by the white arrows in (a) and visible in the height profile. (c) A `side-view' of the structural model, (viewing along the [111] direction, and with the (112) surface normal pointing upwards), shows the division of the structure into trilayer-like slabs stacked in alternating orientations. Planar projections of the two resulting inequivalent surface lattices, labeled I and II, are shown in (d) and (e) respectively.}
\end{figure}

We determine the precise position of the cleavage planes along the (112) normal by inspecting the step heights of the step-terrace morphology along the dashed line in Fig. 2(a). The height profile along this line is shown in Fig. 2(b). The predominant step-height, around 7.2 \AA, corresponds to twice the As-As interlayer spacing parallel to the (112) normal. Figure 2(c) shows a side-view along the [111] direction of the crystal stacking. It can be seen that the basic stacking units are formed by layers of As atoms sandwiched on each side by Cd layers. Each of the observed topographic steps corresponds to a pair of these `trilayer sandwich' stacking units. The pairing is due to the inequivalence of the planes separating the stacking units, which we label as cleavage planes I and II in Fig. 2(c). [And since for each type of cleavage plane, the resulting pair of cleaved surfaces on either side are structurally equivalent with one another, these labels carry over from the cleavage planes to the resulting surface: Cleavage through plane I (or II) always results in a type I (or II) surface.]  As shown in Figs. 2(d) and 2(e), the main structural difference between the two surfaces is that three quarters of the systematic Cd vacancies of the Cd$_{3}$As$_{2}$ structure are located in the Cd atomic layer which terminates surface II. This results in linear channels of Cd vacancies arranged along the [111] direction of the conventional unit cell, indicated by red circles in Fig. 2(c).

In fact cleavage at both planes I and II is observed to occur. In Fig. 2(a), small sub-steps (marked with white arrows) can be seen protruding from the major steps, and have a height corresponding to half the major step height, or one trilayer sandwich, as is shown in the topographic line profile shown in  Fig. 2(b). This indicates that both cleavage planes I and II are observable, but that one is strongly energetically preferred over the other. Here the question remains: Which cleaved surface, I or II, is the predominant one? This can be answered by detailed inspection of the observed surface atomic lattice, to which we will return below. 

At this stage, we conclude that the step heights of the observed step-terrace morphology are consistent with recent descriptions of the Cd$_{3}$As$_{2}$ structure, and that the cleavage planes are located between adjacent Cd atomic layers, which would suggest a Cd terminated surface rather than the As surface previously suggested \cite{Jeon2014}. A Cd terminated surface is in fact more in line with previously reported x-ray photoelectron spectroscopy measurements, which have shown Cd 4\textit{d} doublets dominating in intensity over As 3\textit{d} doublets \cite{Liu2014a,Sankar2015}. (The Cd 4\textit{d} and As 3\textit{d} levels have approximately the same photoionization cross-sections in the low photon energy range \cite{Yeh1993}.)

%\subsection{Interpretation of atomically resolved STM topographs}

Figure 3 shows high-resolution STM topographs of the cleaved (112) surface. The expected pseudohexagonal atomic surface lattice and the previously reported $2 \times 2$ reconstruction are observed. The most outstanding feature is a periodic, uniaxial zig-zag pattern, highlighted with dashed white lines in topographs shown in Figs. 3(a) and (b), which are acquired in the same position but using \textit{V} = -0.5 and -0.1 V respectively. The overall periodicity of the surface superlattice can be characterized with a rectangular unit cell (solid white lines), with lattice parameters $\boldsymbol{a^\prime}$ = 17.9 \AA, $\boldsymbol{b^\prime}$ = 15.6 \AA. Large scale topographs show the ubiquity of this pattern over the entire measured surface \cite{Supplement}. These features were not observed in STM measurements reported by Jeon \textit{et al.} \cite{Jeon2014}, and possible reasons for this discrepancy are discussed below.

Fig. 3(c) shows a scheme for reorienting the Cd$_{3}$As$_{2}$ unit cell so that one of its faces corresponds to the rectangular surface lattice detailed in Figs. 3(a) and (b). The relation between the re-oriented and original lattice vectors is expressed by the unitary transformation $[\boldsymbol{a^\prime}, \boldsymbol{b^\prime}, \boldsymbol{c^\prime}] = [\boldsymbol{a}, \boldsymbol{b}, \boldsymbol{c}]U$, where

\[
U = 
  \begin{bmatrix}
    1 & 0.5 & 0.5\\
    -1 & 0.5 & 0.5\\
    0 & -0.5 & 0.5
  \end{bmatrix}.
\]

This alternative bulk unit cell now forms the basis of our slab model for \textit{ab initio} calculations, and unlike the unit cell shown in Fig. 1(a), allows the construction of a suitable slab with a choice of (112) surface terminations facing the vacuum. Slab models terminated by surfaces I and II were constructed with a thickness of eight `trilayer sandwich' stacking units \cite{Supplement}. Structural optimization resulted in a small deviation from bulk-like terminations (i.e., terminations without any reconstruction) as shown in Figs. 3(d) and 3(e). A simple bulk-like termination for either surface I or II would result in either of the Cd surface lattices as shown in Figs. 2(d) and (e). Instead, in both cases surface relaxation results in a subsidence of the surface Cd lattice down to around the same level as the second (As) layer. Despite this intermingling of the surface Cd and As lattices, the number-density of observed atoms in STM images is not increased compared to that expected for a bulk-like termination. This may be because the density of states (DOS) is strongly dominated either by the Cd 5\textit{s} states (conduction band) or by the As 4\textit{p} states (valence band) \cite{Wang2013}, so that one lattice, either the Cd or As lattice (see below), provides the overwhelming contribution to the tunnel current and topography image. These effects greatly complicate the interpretation of STM images, since it is no longer guaranteed that surfaces I and II can be identified simply by their patterns of  surface lattice vacancies. 

\begin{figure} [h]
\centering
\includegraphics[scale=1]{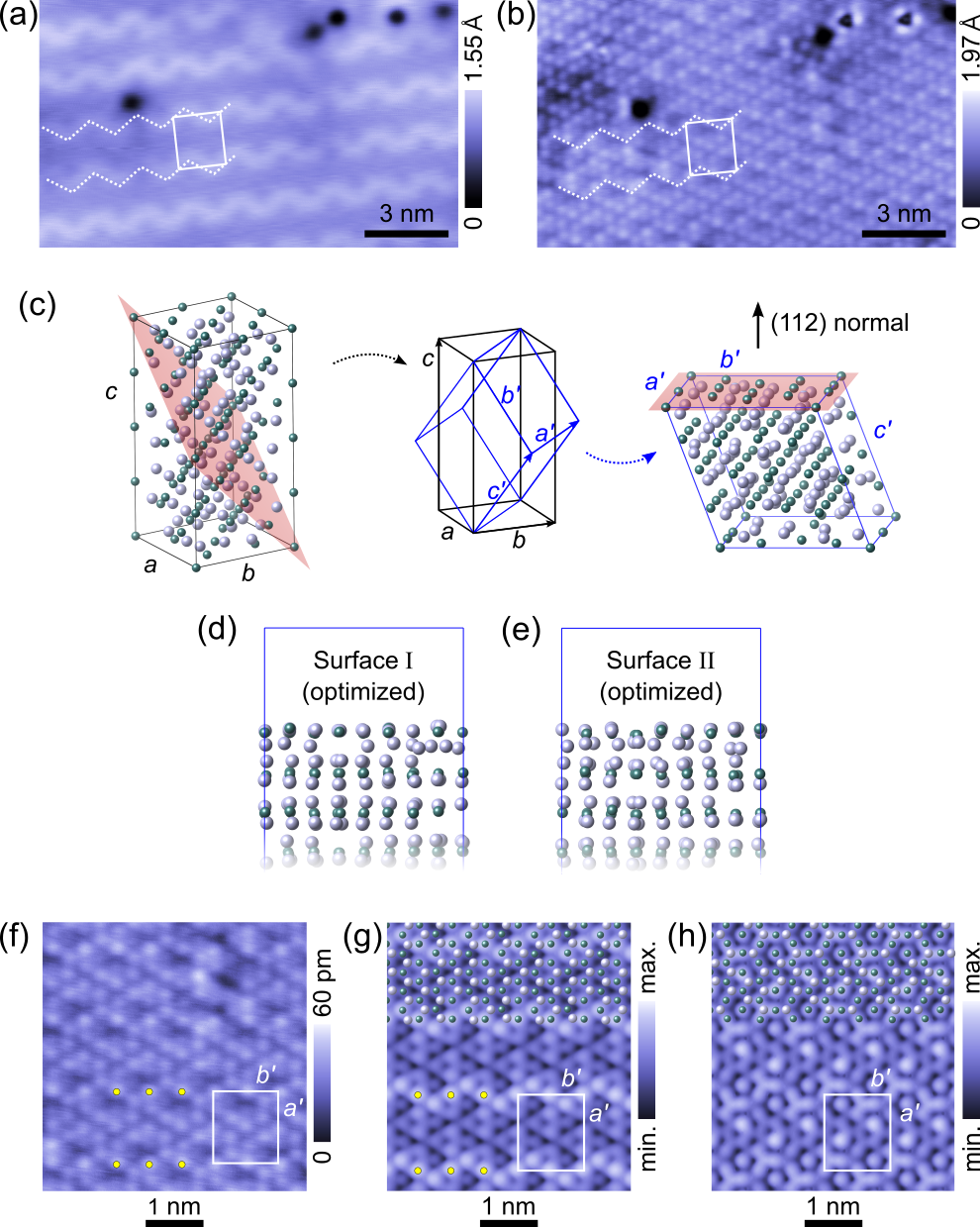}
\caption{\label{fig:3} Interpretation of atomically resolved topography. Zoom-in STM topographs taken using \textit{V} = -0.5 V and -0.1 V, with \textit{I} = 1 nA, are shown in (a) and (b) respectively. Both images show a zig-zag-like superstructure, which can be characterized by a 2-D unit cell shown as a white rectangle in each image. A scheme for setting up a re-oriented unit cell suitable for slab-model calculations is shown in (c). The rectangular upper face of the new unit cell represents the 2-D unit cell identified in (a) and (b). Side-views of the resulting structurally optimized slab-models for surfaces I and II are shown in (d) and (e) respectively. A zoom-in view of the measured STM topography, (f), and the simulated STM images for surfaces I and II, shown in (g) and (h) respectively, are given for comparison.}
\end{figure}

Having obtained optimized structures for the two candidate surfaces, we calculate the partial charge density distribution in order to obtain simulated STM topographs. In order for the simulated and experimental images to be comparable, these partial charge density plots should be obtained by integrating the local density of states (LDOS) between the Fermi level and the energy corresponding to the tip voltage. However, this is not as straightforward as usual: As has been consistently reported in ARPES investigations \cite{Neupane2014,Borisenko2014,Sankar2015} and scanning tunneling spectroscopy measurements \cite{Jeon2014,Sankar2015}, the DPs are found $\sim$200 meV below the Fermi level \cite{Supplement}. This is at odds with the expectation, based on the charge neutrality condition \cite{Wang2013}, that the Dirac nodes reside at or near the Fermi level, as is also predicted using \textit{ab initio} calculations \cite{Wang2013,Neupane2014,Ali2014}. This n-type shift has been speculated to result from donor impurities, with a prime candidate being As vacancies \cite{Spitzer1966}. For the STM image shown in Fig. 3(b), taken at -0.1 eV, the bias is at ($E_{DP}$ + 100) meV. The LDOS integral to obtain the closest corresponding partial charge density map should then be made between $E_{F}$ = ($E_{DP}$ + 200) meV, and ($E_{DP}$ + 100) meV.

Simulated STM maps for candidate surfaces I and II were obtained from calculated partial charge density distributions by considering the Tersoff-Hamann model \cite{TersoffHamann} and using a tip-sample distance of $\sim$4 \AA\enspace \cite{Supplement}. These maps are shown in Figs. 3(g) and 3(h) respectively, alongside a zoom-in STM topography image in Fig. 3(f). In each simulated map, the 2-D projected atom positions for the uppermost layer of both Cd and As are illustrated. Surprisingly, though the cleavage plane was located between two adjacent Cd layers, and the topmost Cd and As layers reside at about the same height, the dominant contribution to the observable surface lattice comes from As, possibly due to a dominance of As 4\textit{p} orbitals in the DOS.

Both of the simulated STM images successfully reproduce the expected rectangular periodicity of the observed superstructure, and an additional periodicity which may correspond to the $2 \times 2$ reconstruction observed by Jeon \textit{et al.} Despite the challenges detailed above, we attempt to match some key features of the simulated STM images with features in the measured STM topograph. Most obviously, the simulated STM for surface I exhibits a pronounced zig-zag pattern, which matches the zig-zag appearing in the measured topographs shown in Figs. 3(a) and (b). The center-line of the zig-zag passes through a line of protrusions, several of which are marked with yellow dots in Fig. 3(g). We find corresponding atoms in the measured topograph [also marked with yellow dots in (f)], which also lie along the center-line of the zig-zag. In both the measured image and the image simulated for surface I, this line of atoms possesses the maximum topographic height. No obviously comparable zig-zag, nor any corresponding set of high protrusions can be found in the simulated STM map for surface II [Fig. 3(h)]. Based on these arguments, we identify the surface observed in measurements as surface I.

%\section{Discussion and Conclusions}

Our initial results, showing that the (112) cleavage plane is located between two adjacent Cd atomic layers, but that the terminating layer is a relaxed mixture of Cd and As atoms, clarifies upon the interpretation given by Jeon \textit{et al.} It was previously thought that the observation of an apparently complete atomic surface lattice (without obvious systematic vacancies) indicated the full \textit{hcp} planar lattice expected for a As terminated (112) surface \cite{Jeon2014}. While the corrugations observed in STM topographs likely correspond to the uppermost As layer, they do not directly reflect the actual surface structure. 

A more concerning difference is the absence in previous work of the surface super-structure described here. This might be explained in a number of ways: i) a difference in samples, with the samples used by Jeon \textit{et al.} possessing a structural phase different from that described in current models \cite{Ali2014,Sankar2015}. ii) a difference in the STM tip condition. In any given STM experiment, the tip apex has its own unknown orbital character, which determines the details of the tunneling transition matrix between the tip and sample, introducing unknown selection effects into topography maps. It is not impossible that different tip conditions may explain the absence in previous reports of the observed super-structure.

In our main result, we provide a microscopic description of a periodic surface superstructure caused by a pattern of systematic Cd vacancies, the symmetry of which underpins the DSM phase in Cd$_{3}$As$_{2}$. This reconciles the status of real-space microscopic observations with the established structure determined using x-ray diffraction (XRD), and with observations of the broad range of physical phenomena thought to accompany it. It also shows that the systematic ordering of vacancies at the surface survives the cleavage process. This surface characterization should help in the understanding of surface and interface phenomena which might be sensitively termination dependent, such as surface Fermi arcs and other possible trivial or non-trivial surface states, or possible unconventional superconductivity at heterostructure interfaces.

\section{Acknowledgements}

The authors thank W.-L. Lee for stimulating discussions. This work was supported in part by the Ministry of Science and Technology and National Science Council of Taiwan through the following grants: MOST 104-2119-M-002-029, and NSC 101-2119-M-002-007.

\end{document}

\title{Supplemental Material: Observation of surface superstructure induced by systematic vacancies in the topological Dirac semimetal Cd$_{3}$As$_{2}$}

\author{Christopher J. Butler$^{1,}$\footnote{These authors contributed equally to this work}$^{,}$\footnote{cjbutler@ntu.edu.tw}}
\author{Yi Tseng$^{1,}$\footnotemark[1]}
\author{Cheng-Rong Hsing$^{2}$}
\author{Yu-Mi Wu$^{1}$}
\author{Raman Sankar$^{3,4}$}
\author{Mei-Fang Wang$^{1}$}
\author{Ching-Ming Wei$^{2,}$\footnote{cmw@phys.sinica.edu.tw}}
\author{Fang-Cheng Chou$^{4,5,6}$}
\author{Minn-Tsong Lin$^{1,2,7,}$\footnote{mtlin@phys.ntu.edu.tw}}

\affiliation{$^{1}$Department of Physics, National Taiwan University, Taipei 10617, Taiwan}
\affiliation{$^{2}$Institute of Atomic and Molecular Sciences, Academia Sinica, Taipei 10617, Taiwan}
\affiliation{$^{3}$Institute of Physics, Academia Sinica, Taipei 11529, Taiwan}
\affiliation{$^{4}$Center for Condensed Matter Sciences, National Taiwan University, Taipei 10617, Taiwan}
\affiliation{$^{5}$National Synchrotron Radiation Research Center, Hsinchu 30076, Taiwan}
\affiliation{$^{6}$Taiwan Consortium of Emergent Crystalline Materials (TCECM), Ministry of Science and Technology, Taipei 10622, Taiwan}
\affiliation{$^{7}$Research Center for Applied Sciences, Academia Sinica, Taipei 11529, Taiwan}

\maketitle\newpage

\begin{figure}
\centering
\includegraphics[scale=1.25]{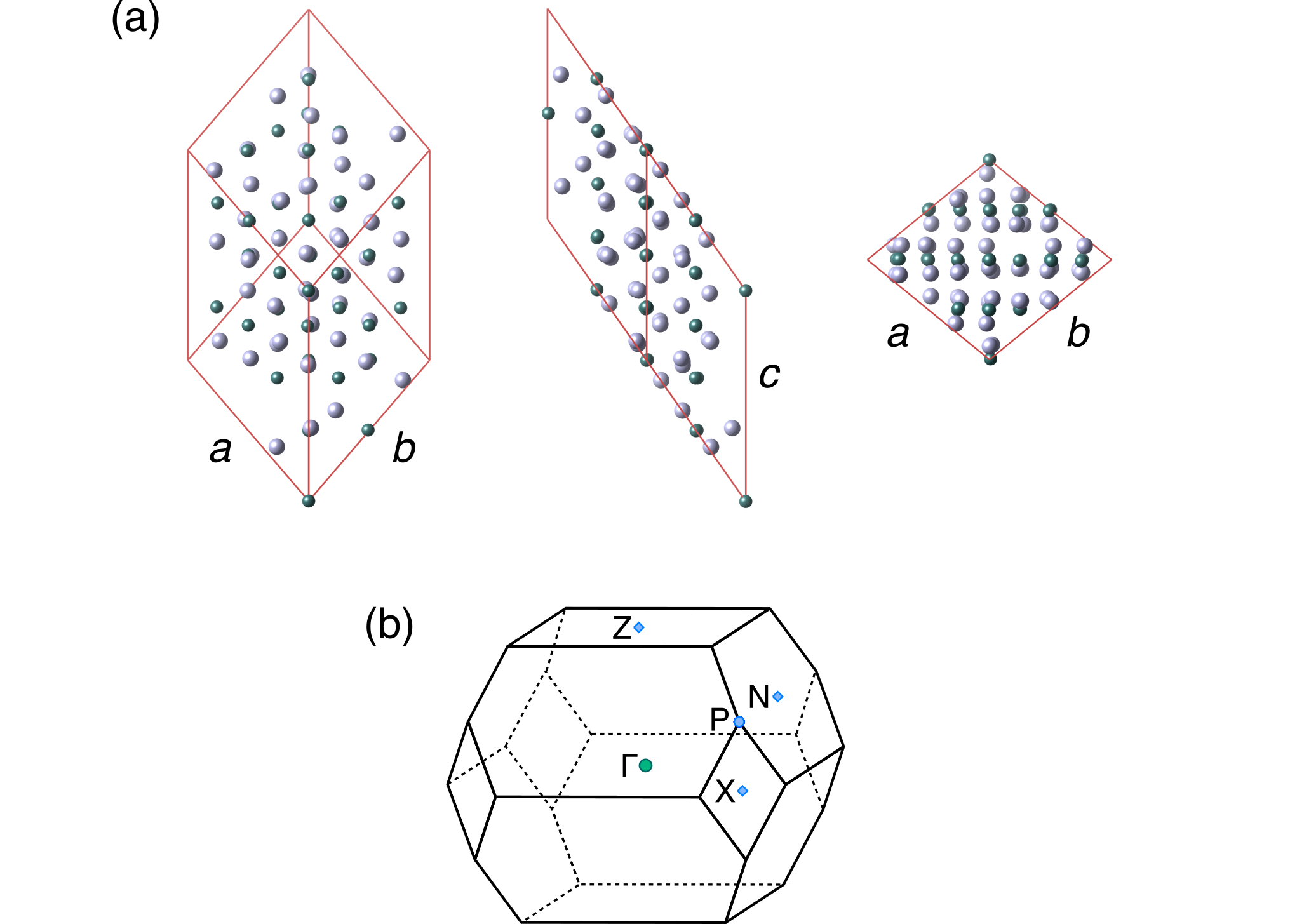}
\caption{\label{fig:1} (a) The primitive unit cell of Cd$_{3}$As$_{2}$, containing 80 atoms, and (b) the corresponding BZ, illustrating the high-symmetry points between which we plot the bulk band diagrams shown in Fig. 1.}
\end{figure}

\newpage

\begin{figure}
\centering
\includegraphics[scale=1.25]{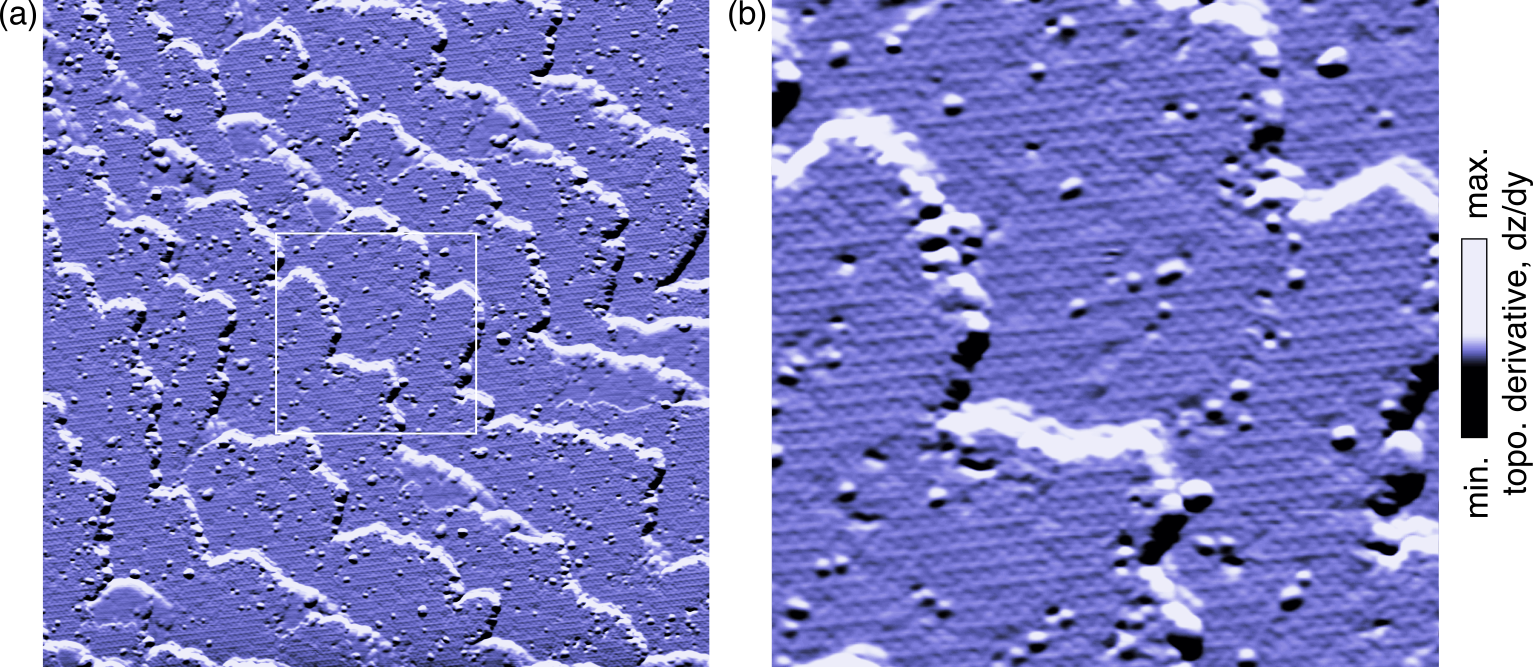}
\caption{\label{fig:2} (a) A typical large scale topographic derivative (dz/dy) map. On close inspection, the zig-zag surface superstructure is seen to be ubiquitous over the entire observed surface. (b) For clarity, a zoom-in image of the area indicated by the white square in panel (a). Both images were acquired using V = 2.0 V, I = 0.1 nA.}
\end{figure}

\newpage

\begin{figure}
\centering
\includegraphics[scale=1.25]{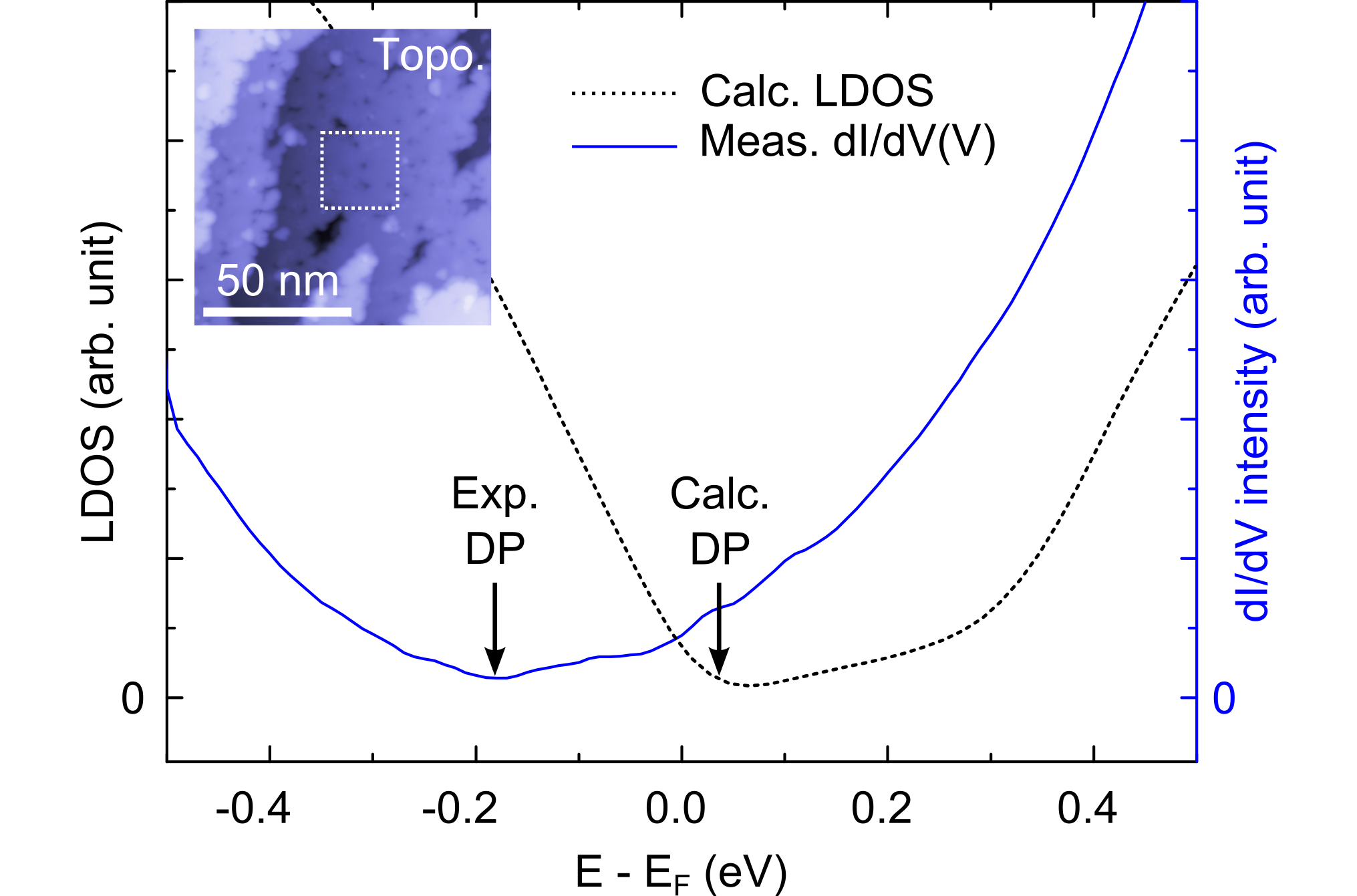}
\caption{\label{fig:3} Typical STS curve (blue) clearly showing a conductance minimum around 200 meV below the Fermi level, a result which is consistent with those obtained previously using ARPES and STS (for example, see Jeon \textit{et al.}, Nat. Mater. \textbf{13,} 4023 (2014)). For comparison, a DOS curve calculated using the slab model for surface I (described in the main text) is also shown, with the DOS minimum slightly above the Fermi level corresponding approximately to the energy of the DPs. DOS curves were calculated as described in the main text, sampling using a 6$\times$6$\times$1 k-point grid, and smoothing using the Methfessel-Paxton scheme (see M. Methfessel, and A. T. Paxton, Phys. Rev. B \textbf{40,} 3616-3621 (1989)), rather than the more usual `tetrahedron method', which proved too computationally intensive for the system at hand.}
\end{figure}

\newpage

\begin{figure}
\centering
\includegraphics[scale=1.25]{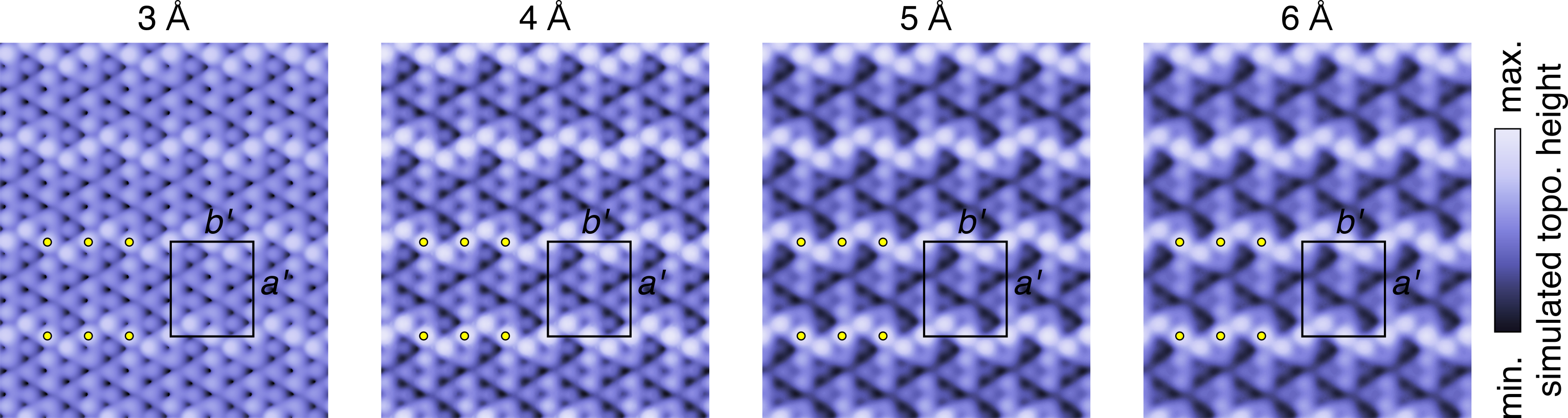}
\caption{\label{fig:4} Simulated STM topography for Cd$_{3}$As$_{2}$(112) surface I, corresponding to various tip-sample distances, with the approximate distance marked above each panel. The qualitative features which can be matched to those seen in the atomically resolved STM image in Fig. 3(f) are present in all simulated images: The unit cell describing the vacancy-induced surface superstructure is marked by a white rectangle, and a number of the highest atomic protrusions are marked with yellow dots, for comparison with Fig. 3. The colormap is applied separately to each image.}
\end{figure}